\documentclass[hyper]{JHEP3}
\usepackage{amsmath,amssymb,amsfonts,eepic,epsfig,psfrag}
\author{V.A.~Fateev$^{1,2}$ and A.V.~Litvinov$^{1,3}$\\
$^1$~Landau Institute for Theoretical Physics, 142432, Russia, Moscow Region, Chernogolovka, acad. Semenov prosp., 1a.\\
$^2$~Laboratoire de Physique Th\'eorique et Astroparticules, UMR5207 CNRS-UM2, Universit\'e
Montpellier~II, Pl.~E.~Bataillon, 34095 Montpellier, France\\
$^3$~NHETC, Department of Physics and Astronomy, Rutgers University,\\ 136 Frelinghuysen Road, Piscataway, NJ 08855-0849, U.S.A.}
\abstract{In these notes we consider relation between conformal blocks and the Nekrasov partition function of certain $\mathcal{N}=2$ SYM theories proposed recently by Alday, Gaiotto and Tachikawa. We concentrate on $\mathcal{N}=2^{*}$ theory, which is the simplest example of AGT relation.}
\title{On AGT conjecture}
\preprint{RUNHETC-2009-22}
\begin{document}
In these notes we consider relation between four-dimensional $\mathcal{N}=2$ supersymmetric gauge theories and Liouville field theory (LFT) proposed recently in \cite{Alday:2009aq}. In particular, it was conjectured in \cite{Alday:2009aq} that the Nekrasov partition function \cite{Nekrasov:2002qd} in certain $\mathcal{N}=2$ supersymmetric gauge theories coinsides up to some trivial factors with the conformal blocks in LFT. This conjecture was checked in particular cases in \cite{Alday:2009aq} and lately in  \cite{Mironov:2009qt,Mironov:2009qn,Marshakov:2009kj}. For simplicity we consider $\mathcal{N}=2^{*}$ $U(2)$ SYM theory, i.e. $\mathcal{N}=4$ theory deformed by the mass to the adjoint hypermultiplet. We show that the Nekrasov partition function in such a theory is related with the one-point conformal block of the primary field in LFT on a torus. 

In section \ref{One-point-block} we remind the definition of the one-point conformal block on a torus and derive recursive formula for it. In section \ref{Nekrasov-partition-funtion} we consider integral representation for the Nekrasov partition function in $\mathcal{N}=2^{*}$ theory and show that it satisfies exactly the same recursive formula. We consider also asymptotic of the partition function at large values of the VEV of the adjoint scalar and find precise correspondence between these two objects. In section \ref{SW} we consider the limit $\varepsilon_1,\varepsilon_2\rightarrow0$ of the partition function in which the Seiberg-Witten curve is recovered and obtain WKB like parametric formula for the instanton part of the prepotential. In appendices we collect some proofs as well as some useful formulae.

Our approach can also be applied to the $\mathcal{N}=2$ theory with four fundamental flavors which is related with four-point conformal block in LFT on a sphere \cite{Alday:2009aq} in more or less straitforward way. We think that our approach can also be applied to the AGT conjecture for higher rank gauge groups \cite{Wyllard:2009hg,Mironov:2009by} as well as to more complicated quiver gauge theories \cite{Alday:2009aq,Gaiotto:2009we,Fucito:2004gi}.
\section{Recursive formula for the one-point toric conformal block}\label{One-point-block}
First of all we want to stress that we don't really need the Liouville field theory because we will consider only conformal blocks which are objects completely fixed by the conformal invariance \cite{Belavin:1984vu}.
We will use Liouville parameterization for the central charge
\begin{equation}
     c=1+6Q^2,\quad\text{where}\quad Q=b+\frac{1}{b},
\end{equation}
and for the dimensions of the primary fields
\begin{equation}
      \Delta(\alpha)=\alpha(Q-\alpha).
\end{equation}
In these notations degenerate representations of the Virasoro algebra are labeled by
\begin{equation}\label{alpha-degenerate}
    \alpha_{m,n}=-\frac{mb}{2}-\frac{n}{2b}\quad\text{with}\quad m,n=0,1,2\dots
\end{equation}
and we denote
\begin{equation}\label{Delta-degenerate}
   \Delta_{m,n}\overset{\text{def}}{=}\Delta(\alpha_{m-1,n-1})
\end{equation}

One-point correlation function on a torus with modulus $\tau$ ($q=e^{2i\pi\tau}$) is defined by
\begin{equation}
    \langle V_{\alpha}\rangle_{\tau}=
    \text{Tr}\left(q^{L_0-\frac{c}{24}}\bar{q}^{\bar{L}_0-\frac{c}{24}}V_{\alpha}(0)\right)
    =
    \sum_{\{\Delta\}} C_{\Delta_{\alpha}\Delta}^{\Delta}
    \,|q|^{2\Delta-\frac{c}{12}}\left|\mathcal{F}_{\alpha}^{(\Delta)}(q)\right|^2,
\end{equation}
where one-point conformal block $\mathcal{F}_{\alpha}^{(\Delta)}(q)$ is defined as the contribution to the trace of the conformal family with conformal dimension $\Delta$
\begin{multline}\label{one-point-block-expansion}
     \mathcal{F}_{\alpha}^{(\Delta)}(q)\overset{\text{def}}{=}\frac{1}{\langle\Delta|V_{\alpha}|\Delta\rangle}
     \left(\langle\Delta|V_{\alpha}|\Delta\rangle+
     \frac{\langle\Delta|L_{1}V_{\alpha}L_{-1}|\Delta\rangle}
     {\langle\Delta|L_{1}L_{-1}|\Delta\rangle}q+\dots\right)=\\=
     1+\frac{2\Delta+\Delta^2(\alpha)-\Delta(\alpha)}{2\Delta}q+\dots
\end{multline}
Structure constants of the operator algebra
\begin{equation}
     C_{\Delta_{\alpha}\Delta}^{\Delta}=\langle\Delta|V_{\alpha}|\Delta\rangle,
\end{equation}
as well as the set of intermediate dimesions $\{\Delta\}$ are details of the theory, but the conformal block \eqref{one-point-block-expansion} is universal object. Arbitrary matrix element in \eqref{one-point-block-expansion}
\begin{equation}
    \langle\Delta|L_{l_1}\dots L_{l_m}V_{\alpha}L_{-k_1}\dots L_{-k_n}|\Delta\rangle=\lim_{z\rightarrow1}\,
    \langle\Delta|L_{l_1}\dots L_{l_m}V_{\alpha}(z)L_{-k_1}\dots L_{-k_n}|\Delta\rangle,
\end{equation}
can be obtained from the three-point function of primary fields $\langle\Delta|V_{\alpha}(z)|\Delta\rangle\sim z^{-\Delta(\alpha)}$ using commutation relations
\begin{equation}
     [L_n,V_{\alpha}(z)]=z^{n+1}\partial V_{\alpha}(z)+\Delta(\alpha)(n+1)z^{n}V_{\alpha}(z).
\end{equation}
This purely algebraic exercise becomes tedious for higher levels and here we would like to consider alternative approach first suggested in \cite{Zamolodchikov:1985ie} for the case of four-point conformal block on a sphere\footnote{Actually, our case (one-point toric conformal block) is much more simple than the general four-point spheric conformal block considered in \cite{Zamolodchikov:1985ie}. As it follows from the results of the paper \cite{Fateev:2009me} the former is the particular case of it. Reccursion formula \eqref{One-point-block-recursion} derived in this paper was considered recently in \cite{Poghossian:2009mk} and proved in \cite{Hadasz:2009db}.}. 
We notice that the conformal block $\mathcal{F}_{\alpha}^{(\Delta)}(q)$ as a function of the intermediate dimension $\Delta$ has simple poles at
\begin{equation}
      \Delta=\Delta_{m,n},
\end{equation}
where $\Delta_{m,n}$ is the conformal dimension of degenerate field \eqref{Delta-degenerate}.
To see this let us consider the state $|\chi\rangle$ on the level $N=mn$ which has the limit at $\Delta\rightarrow\Delta_{m,n}$
\begin{equation}
    |\chi\rangle\rightarrow|\chi_{m,n}\rangle,
\end{equation}
where $|\chi_{m,n}\rangle$ is the null-vector at the level $mn$ which we normalize as
\begin{equation}
      |\chi_{m,n}\rangle=\bigl(L_{-1}^{mn}+\dots\bigr)|\Delta_{m,n}\rangle.
\end{equation}
By definition, this vector is killed by the positive part of the Virasoro algebra and has conformal dimension $\Delta_{m,n}+mn=\Delta_{m,-n}$ \cite{Belavin:1984vu}. Moreover, it has a vanishing norm
\begin{equation}
     \langle\chi|\chi\rangle=(\Delta-\Delta_{m,n})r_{m,n}+\dots
\end{equation}
where \cite{Zamolodchikov:1985ie,Zamolodchikov:2003yb}
\addtocounter{equation}{-1}
\begin{subequations}
\begin{equation}
    r_{m,n}=2\prod_{k=1-m}^m\prod_{l=1-n}^n(kb+lb^{-1}),
\end{equation}
\end{subequations}
with $(k,l)\neq(0,0)$, $(k,l)\neq(m,n)$ and is orthonormal to other states. Furthermore, any state at the level $N>mn$ of the form $L|\chi\rangle$ where $L$ is some combination of the Virasoro generators of total degree $N-mn$ will have zero norm at  $\Delta\rightarrow\Delta_{m,n}$ and hence will contribute to the singular part of the conformal block $\mathcal{F}_{\alpha}^{(\Delta)}(q)$. This contribution has the form\footnote{In order to obtain \eqref{trace-element-pole} it is convenient to use that in the limit $\Delta\rightarrow\Delta_{m,n}$ 
\begin{equation*}
    \langle\chi|L^{+}V_{\alpha}L|\chi\rangle= 
    \frac{\langle\Delta_{m,-n}|L^{+}V_{\alpha}L|\Delta_{m,-n}\rangle}
    {\langle\Delta_{m,-n}|V_{\alpha}|\Delta_{m,-n}\rangle}\,\langle\chi|V_{\alpha}|\chi\rangle+\dots.
\end{equation*}}
\begin{multline}\label{trace-element-pole}
    \frac{1}{\langle\Delta|V_{\alpha}|\Delta\rangle} 
    \frac{\langle\chi|L^{+}V_{\alpha}L|\chi\rangle}{\langle\chi|L^{+}L|\chi\rangle}=
    \frac{R_{m,n}(\alpha)}{(\Delta-\Delta_{m,n})}
    \times\\\times
    \frac{1}{\langle\Delta_{m,-n}|V_{\alpha}|\Delta_{m,-n}\rangle} 
    \frac{\langle\Delta_{m,-n}|L^{+}V_{\alpha}L|\Delta_{m,-n}\rangle}
    {\langle\Delta_{m,-n}|L^{+}L|\Delta_{m,-n}\rangle}+\dots
\end{multline}
where by definition 
\begin{equation}\label{Rmn}
    R_{m,n}(\alpha)=r_{m,n}^{-1}\,
    \frac{\langle\chi_{m,n}|V_{\alpha}|\chi_{m,n}\rangle}{\langle\Delta_{m,n}|V_{\alpha}|\Delta_{m,n}\rangle}.   
\end{equation}
Matrix element in \eqref{Rmn} can be extracted from ref \cite{Zamolodchikov:2003yb}
\addtocounter{equation}{-1}
\begin{subequations}
\begin{equation}
    \frac{\langle\chi_{m,n}|V_{\alpha}|\chi_{m,n}\rangle}{\langle\Delta_{m,n}|V_{\alpha}|\Delta_{m,n}\rangle}=
    \prod_{k,l}\left(\frac{Q}{2}-\alpha-\alpha_{k,l}\right)
\end{equation}
where
\begin{equation*}
    \begin{aligned}
       &k=-(2m-1),-(2m-3),\dots,(2m-3),(2m-1),\\
       &l=-(2n-1),-(2n-3),\dots,(2n-3),(2n-1),
    \end{aligned}
\end{equation*}
\end{subequations}
and $\alpha_{k,l}$ are given by \eqref{alpha-degenerate}.
From eq \eqref{trace-element-pole} it is clear that the residue of the  conformal block $\mathcal{F}_{\alpha}^{(\Delta)}(q)$ at $\Delta=\Delta_{m,n}$ is proportional to $\mathcal{F}_{\alpha}^{(\Delta_{m,-n})}(q)$
\begin{equation}\label{Elliptic-block-res}
     \text{Res}\,\mathcal{F}_{\alpha}^{(\Delta)}(q)\Biggl|_{\Delta=\Delta_{m,n}}\hspace*{-15pt}=
     q^{mn}R_{m,n}(\alpha)\,\mathcal{F}_{\alpha}^{(\Delta_{m,-n})}(q).
\end{equation}
Using \eqref{Elliptic-block-res} one can write
\begin{equation}\label{Elliptic-block-res-sum}
       \mathcal{F}_{\alpha}^{(\Delta)}(q)=\sum_{m,n}q^{mn}\frac{R_{m,n}(\alpha)}{\Delta-\Delta_{m,n}}\,
       \mathcal{F}_{\alpha}^{(\Delta_{m,-n})}(q)+f_{\alpha}^{(\Delta)}(q),
\end{equation}
where function $f_{\alpha}^{(\Delta)}(q)$ corresponds to the limit of the one-point conformal block at $\Delta\rightarrow\infty$ with $\alpha$ being kept fixed. This asymptotic can be obtained from the fact that for any state $L|\Delta\rangle$ one has
\begin{equation}
        \frac{\langle\Delta|L^{+}V_{\alpha}L|\Delta\rangle}
        {\langle\Delta|V_{\alpha}|\Delta\rangle\langle\Delta|L^{+}L|\Delta\rangle}=1+
        O\left(\frac{1}{\Delta}\right)\qquad\text{at}\qquad\Delta\rightarrow\infty.
\end{equation}
So that in the limit $\Delta\rightarrow\infty$ the contribution of each state is just $1$ and the conformal block is equal to the character which is given by
\begin{equation}
       \mathcal{F}_{\alpha}^{(\Delta)}(q)
       \xrightarrow[\Delta\rightarrow\infty]{}\frac{q^{\frac{1}{24}}}{\eta(\tau)}=1+q+2q^2+\dots,
\end{equation}
here $\eta(\tau)$ is Dedekind eta function. If we assume that the poles in \eqref{Elliptic-block-res-sum} are the only singularities of the function $\mathcal{F}_{\alpha}^{(\Delta)}(q)$ then
\begin{equation}
    f_{\alpha}^{(\Delta)}(q)=\frac{q^{\frac{1}{24}}}{\eta(\tau)}.
\end{equation}
Defining
\begin{equation}\label{F-H-definition}
        \mathcal{F}_{\alpha}^{(\Delta)}(q)=\frac{q^{\frac{1}{24}}}{\eta(\tau)}\,\mathcal{H}_{\alpha}^{(\Delta)}(q)
\end{equation}
we arrive that function $\mathcal{H}_{\alpha}^{(\Delta)}(q)$ satisfies recursive relation
\begin{equation}\label{One-point-block-recursion}
     \mathcal{H}_{\alpha}^{(\Delta)}(q)=1+\sum_{m,n}q^{mn}\frac{R_{m,n}(\alpha)}{\Delta-\Delta_{m,n}}\,
       \mathcal{H}_{\alpha}^{(\Delta_{m,-n})}(q).
\end{equation}
Relation \eqref{One-point-block-recursion} for $\mathcal{H}_{\alpha}^{(\Delta)}(q)$ is happen to be very effective for calculation of its expansion in power series of $q$. Namely, let us represent $\mathcal{H}_{\alpha}^{(\Delta)}(q)$ as
\begin{equation}\label{Hblock-series}
    \mathcal{H}_{\alpha}^{(\Delta)}(q)=1+\sum_{L=1}^{\infty}H_L(\Delta)q^L. 
\end{equation}
Relation \eqref{One-point-block-recursion} leads to recursive algorithm for the coefficients $H_L(\Delta)$ (we define here for convenience $H_0(\Delta)=1$)
\begin{equation}
   H_L(\Delta)=\sum_{mn\leq L}\frac{R_{m,n}(\alpha)}{\Delta-\Delta_{m,n}}H_{L-mn}(\Delta_{m,-n}).
\end{equation}
\section{$\mathcal{N}=2^{*}$ $U(2)$ Nekrasov instanton partition function}\label{Nekrasov-partition-funtion}
We consider $\mathcal{N}=2$ $U(2)$ SYM theory with matter in adjoint representation deformed by the mass term. The instanton part of the partition function is given by \cite{Nekrasov:2002qd}
\begin{equation}\label{Partition-function-N=2*-def}
   Z_{\text{inst}}^{\mathcal{N}=2^{*}}(\varepsilon_1,\varepsilon_2,m,\vec{a})
   =1+\sum_{k=1}^{\infty}q^{k}\mathfrak{Z}_k,
\end{equation} 
where $\mathfrak{Z}_N$ is given by the $N-$dimensional integral (see (3.25) in \cite{Nekrasov:2002qd}). Here $m$ is the mass of the matter multiplet, $\vec{a}=(a_1,a_2)$ is VEV of the complex scalars and $\varepsilon_1$, $\varepsilon_2$ are the deformation parameters of $\Omega$ background (see \cite{Nekrasov:2002qd}). Instanton parameter $q$ in \eqref{Partition-function-N=2*-def} is given by
\begin{equation}
   q=e^{2i\pi\tau},\quad\text{where}\quad \tau=\frac{4i\pi}{g^2}+\frac{\theta}{2\pi}.
\end{equation}
Let us modify the notations which will be more suitable in our case
\begin{equation}\label{parms-rescale}
   \begin{aligned}
       &\vec{a}=\hbar\vec{P},\qquad
       &&m=\hbar\alpha,\\
       &\varepsilon_1=\hbar b,\qquad
       &&\varepsilon_2=\frac{\hbar}{b}.
   \end{aligned}
\end{equation}
Total mass scale $\hbar$ in $\mathfrak{Z}_N$ disappears and we can rewrite Nekrasov integral \cite{Nekrasov:2002qd} as
\begin{multline}\label{Zn-integral}
  \mathfrak{Z}_N=
  \frac{1}{N!}\left(\frac{Q(b-\alpha)(b^{-1}-\alpha)}{2\pi\mathrm{i}\alpha(Q-\alpha)}\right)^N
  \oint\limits_{\mathcal{C}_1}\dots\oint\limits_{\mathcal{C}_N}
  \prod_{k=1}^N\frac{\mathcal{P}(x_k+\alpha)\mathcal{P}(x_k+Q-\alpha)}{\mathcal{P}(x_k)\mathcal{P}(x_k+Q)}
  \times\\\times
  \prod_{i<j}\frac{x_{ij}^2(x_{ij}^2-Q^2)(x_{ij}^2-(b-\alpha)^2)(x_{ij}^2-(b^{-1}-\alpha)^2)}
  {(x_{ij}^2-b^2)(x_{ij}^2-b^{-2})(x_{ij}^2-\alpha^2)(x_{ij}^2-(Q-\alpha)^2)}\,dx_1\dots dx_N,
\end{multline}
where\footnote{In the $U(N)$ case $\mathcal{P}(x)=(x-P_1)\dots(x-P_N)$.} 
\addtocounter{equation}{-1}
\begin{subequations}
\begin{equation}
\mathcal{P}(x)=(x-P_1)(x-P_2) 
\end{equation}
\end{subequations}
and $x_{ij}=x_i-x_j$. The contour $\mathcal{C}_k$ surrounds poles $x_k=P_1$, $x_k=P_2$, $x_k=x_j+b$ and $x_k=x_j+b^{-1}$. Integral \eqref{Zn-integral} can be expressed as a sum over pairs of Young diagrams. Namely, let us denote the situation with no integrals being taken as a pair of empty diagrams
\begin{equation*}
     \vec{Y}_0=(\varnothing,\varnothing).
\end{equation*}
We can arrange all integration variables $x_1,\dots,x_N$ in the following way.
The integral over variable $x_1$ surrounds poles $P_1$ and $P_2$. One has to choose one of them, for example $P_1$. It can be drawn as
\begin{equation*}
    \vec{Y}_1=\left(
   \begin{picture}(30,12)(19,22)
    \Thicklines
    \unitlength 1pt 
    \put(20,20){\line(0,1){10}}
    \put(30,20){\line(0,1){10}}
    \put(20,20){\line(1,0){10}}
    \put(20,30){\line(1,0){10}}
    \put(35,20){\mbox{, $\varnothing$}}
  \end{picture}\right)
\end{equation*}
The integral over variable $x_2$ surrounds more poles. First of all it surrounds $P_2$ (we note that $P_1$ is no longer the pole due to the term $x_{12}^2$ in the numerator in \eqref{Zn-integral}), but also $P_1+b$ and $P_1+b^{-1}$ (these poles come from the terms $x_{12}^2-b^2$ and $x_{12}^2-b^{-2}$ in the denominator in \eqref{Zn-integral}). One has to choose one of the above possibilities. They can be drawn as (we choose the convention to draw shifts in $b$ horizontalally and shifts in $b^{-1}$ vertically)
\begin{equation*}
    \vec{Y}_2=
   \left(
   \begin{picture}(35,12)(19,22)
    \Thicklines
    \unitlength 1pt 
    \put(20,20){\line(0,1){10}}
    \put(30,20){\line(0,1){10}}
    \put(20,20){\line(1,0){10}}
    \put(20,30){\line(1,0){10}}
    \put(35,20){\mbox{,}}
    \put(42,20){\line(0,1){10}}
    \put(52,20){\line(0,1){10}}
    \put(42,20){\line(1,0){10}}
    \put(42,30){\line(1,0){10}}
  \end{picture}\right),\quad
\vec{Y}_2=
   \left(
   \begin{picture}(40,12)(20,22)
    \Thicklines
    \unitlength 1pt 
    \put(20,20){\line(0,1){10}}
    \put(30,20){\line(0,1){10}}
    \put(20,20){\line(1,0){10}}
    \put(20,30){\line(1,0){10}}
    \put(30,20){\line(0,1){10}}
    \put(40,20){\line(0,1){10}}
    \put(30,20){\line(1,0){10}}
    \put(30,30){\line(1,0){10}}
    \put(45,20){\mbox{, $\varnothing$}}
  \end{picture}\right)\quad\text{or}\quad
\vec{Y}_2=
   \left(
   \begin{picture}(30,17)(19,25)
    \Thicklines
    \unitlength 1pt 
    \put(20,20){\line(0,1){10}}
    \put(30,20){\line(0,1){10}}
    \put(20,20){\line(1,0){10}}
    \put(20,30){\line(1,0){10}}
    \put(20,30){\line(0,1){10}}
    \put(30,30){\line(0,1){10}}
    \put(20,30){\line(1,0){10}}
    \put(20,40){\line(1,0){10}}
    \put(35,20){\mbox{, $\varnothing$}}
  \end{picture}\right)\quad\text{respectively.}
\end{equation*}
While integrating over variable $x_3$ and further one has to keep in mind which poles were chosen in the previous steps. This would correspond to drawing the square in appropriate place. It is trivially to see that the resulting "picture" at any step will look like a pair of Young diagrams. It means that diagrams like
\begin{equation*}
    \begin{picture}(30,17)(19,25)
    \Thicklines
    \unitlength 1pt 
    \put(20,20){\line(0,1){10}}
    \put(30,20){\line(0,1){10}}
    \put(20,20){\line(1,0){10}}
    \put(20,30){\line(1,0){10}}
    \put(20,30){\line(0,1){10}}
    \put(30,30){\line(0,1){10}}
    \put(20,30){\line(1,0){10}}
    \put(20,40){\line(1,0){10}}
    \put(30,30){\line(0,1){10}}
    \put(40,30){\line(0,1){10}}
    \put(30,30){\line(1,0){10}}
    \put(30,40){\line(1,0){10}}
  \end{picture}\quad\text{or}\quad
  \begin{picture}(30,17)(19,25)
    \Thicklines
    \unitlength 1pt 
    \put(20,20){\line(0,1){10}}
    \put(30,20){\line(0,1){10}}
    \put(20,20){\line(1,0){10}}
    \put(20,30){\line(1,0){10}}
    \put(30,30){\line(0,1){10}}
    \put(40,30){\line(0,1){10}}
    \put(30,30){\line(1,0){10}}
    \put(30,40){\line(1,0){10}}
    \put(30,20){\line(0,1){10}}
    \put(40,20){\line(0,1){10}}
    \put(30,20){\line(1,0){10}}
    \put(30,30){\line(1,0){10}}
  \end{picture}
\end{equation*}
are forbidden (this is due to the terms $(x_{ij}^2-Q^2)$ in the numerator in \eqref{Zn-integral} which cancell corresponding poles). The resulting answer for the integral \eqref{Zn-integral} can be written as a sum over all possible pairs of Young diagrams with the total number of cells equal to $N$ \cite{Nekrasov:2002qd} (including the cases when one diagram is empty). We note that there are exactly $N!$ ways to obtain the same pair from the integral \eqref{Zn-integral}.
Let $\vec{Y}=(Y_1,Y_2)$ be such pair. Then $\mathfrak{Z}_N$ is given by
\begin{equation}\label{Partition-function-N=2*-redefined}
   \mathfrak{Z}_N
   =\sum_{\vec{Y}}\prod_{i,j=1}^2\prod_{s\in Y_i}
   \frac{(E_{ij}(s)-\alpha)(Q-E_{ij}(s)-\alpha)}
   {E_{ij}(s)(Q-E_{ij}(s))},
\end{equation}
where
\addtocounter{equation}{-1}
\begin{subequations}
\begin{equation}\label{E_ij-definition}
     E_{ij}(s)=P_i-P_j-bH_{Y_j}(s)+b^{-1}(V_{Y_i}(s)+1).
\end{equation}
\end{subequations}
In \eqref{E_ij-definition} $H_Y(s)$ and $V_Y(s)$ are respectively the horizontal and vertical distances from the square $s$ to the edge of the diagram $Y$ (we note, that by definition $s$ in \eqref{Partition-function-N=2*-redefined} always belongs to the diagram $Y_i$). For example, let
\begin{equation*}
    \vec{Y}=(Y_1,Y_2)=\left(
   \begin{picture}(60,25)(20,30)
    \Thicklines
    \unitlength 1pt 
    \put(20,20){\line(0,1){30}}
    \put(30,20){\line(0,1){30}}
    \put(20,20){\line(1,0){10}}
    \put(20,30){\line(1,0){10}}
    \put(20,40){\line(1,0){10}}
    \put(20,50){\line(1,0){10}}
    \put(40,20){\line(0,1){20}}
    \put(30,20){\line(1,0){10}}
    \put(30,30){\line(1,0){10}}
    \put(30,40){\line(1,0){10}}
    \put(50,20){\line(0,1){10}}
    \put(40,20){\line(1,0){10}}
    \put(40,30){\line(1,0){10}}
    \put(55,20){\mbox{,}}
    \put(70,20){\line(0,1){10}}
    \put(80,20){\line(0,1){10}}
    \put(70,20){\line(1,0){10}}
    \put(70,30){\line(1,0){10}}
  \end{picture}\right)
\end{equation*}
It is convenient to think about $s$ as a pair $(k,l)$ with $k,l=1,2,\dots$ Using these notations we have for example
\begin{equation*}
       H_{Y_2}\left((1,1)\right)=0\quad\text{and}\quad V_{Y_1}\left((1,1)\right)=2.
\end{equation*}
We note that $H_{Y}$ and $V_{Y}$ can be also negative. For example,
\begin{equation*}
       H_{Y_2}\left((1,2)\right)=-1\quad\text{and}\quad V_{Y_1}\left((1,2)\right)=1.
\end{equation*}
It is not a big deal to compare few first terms in the expansion \eqref{Partition-function-N=2*-def} with the first few terms in the expansion of the one-point conformal block \eqref{one-point-block-expansion}. It suggests the following identification \cite{Alday:2009aq}
\begin{equation}\label{AGT-relation}
    Z_{\text{inst}}^{\mathcal{N}=2^{*}}(\varepsilon_1,\varepsilon_2,m,\vec{a})=
    \left(\frac{q^{\frac{1}{24}}}{\eta(\tau)}\right)^{1-2\Delta(\alpha)}\mathcal{F}_{\alpha}^{(\Delta)}(q),
\end{equation}
where
\addtocounter{equation}{-1}
\begin{subequations}
\begin{equation}
    \Delta=\frac{Q^2}{4}-P^2,\quad\text{with}\quad P=\frac{P_1-P_2}{2}.
\end{equation}
\end{subequations}

\begin{figure}
\psfrag{a1}{$P_1$}
\psfrag{a2}{$P_2$}
\psfrag{b1}{$P_1-Q$}
\psfrag{b2}{$P_2-Q$}
	\centering
	\includegraphics[width=.6\textwidth]{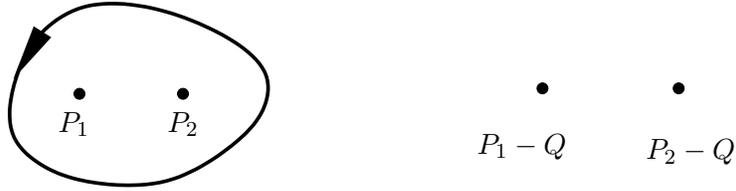}
        \caption{Integration contour in  \protect\eqref{Zn-integral} surrounds poles $P_1$ and $P_2$ 
        while poles $P_1-Q$ and $P_2-Q$ lie outside. When two of them collide (for example $P_1$ and $P_2-Q$) 
        the integral \protect\eqref{Zn-integral} occur a pole.}
	\label{Colliding contour}
\end{figure}

In order to prove relation \eqref{AGT-relation} we consider the structure of singularities of the integral \eqref{Zn-integral}. A singularity can happen when two poles of the integrand collide. For example, if the pole $P_1$ which is inside the contour collide with the pole $P_2-Q$ outside the contour (see fig \ref{Colliding contour}).  
More general pole happens when $P_1+(m-1)b+(n-1)b^{-1}$ collide with $P_2-Q$. In order to archive the pole $P_1+(m-1)b+(n-1)b^{-1}$ one need "draw" the following pair of Young diagrams
\begin{equation*}
     \vec{Y}=(Y_1,Y_2)=\left(
   \begin{picture}(120,50)(0,50)
    \Thicklines
    \unitlength 1pt 
    \put(20,20){\line(0,1){30}}
    \put(30,20){\line(0,1){30}}
    \put(20,20){\line(1,0){10}}
    \put(20,30){\line(1,0){10}}
    \put(20,40){\line(1,0){10}}
    \put(20,50){\line(1,0){10}}
    \put(40,20){\line(0,1){20}}
    \put(30,20){\line(1,0){10}}
    \put(30,30){\line(1,0){10}}
    \put(30,40){\line(1,0){10}}
    \put(50,20){\line(0,1){10}}
    \put(40,20){\line(1,0){10}}
    \put(40,30){\line(1,0){10}}
    \put(60,20){\line(0,1){10}}
    \put(50,20){\line(1,0){10}}
    \put(50,30){\line(1,0){10}}
    \put(60,30){\line(0,1){10}}
    \put(50,30){\line(1,0){10}}
    \put(50,40){\line(1,0){10}}
    \put(50,30){\line(0,1){10}}
    \put(40,30){\line(1,0){10}}
    \put(40,50){\line(1,0){10}}
    \put(60,40){\line(0,1){10}}
    \put(50,40){\line(1,0){10}}
    \put(50,50){\line(1,0){10}}
    \put(50,40){\line(0,1){10}}
    \put(40,40){\line(1,0){10}}
    \put(40,50){\line(1,0){10}}
    \put(40,40){\line(0,1){10}}
    \put(30,40){\line(1,0){10}}
    \put(30,50){\line(1,0){10}}
    
    \put(60,65){\line(0,1){10}}
    \put(50,65){\line(1,0){10}}
    \put(50,75){\line(1,0){10}}
    \put(50,65){\line(0,1){10}}
    \put(40,65){\line(1,0){10}}
    \put(40,75){\line(1,0){10}}
    \put(40,65){\line(0,1){10}}
    \put(30,65){\line(1,0){10}}
    \put(30,75){\line(1,0){10}}
    \put(30,65){\line(0,1){10}}
    \put(20,65){\line(1,0){10}}
    \put(20,75){\line(1,0){10}}
    \put(20,65){\line(0,1){10}}
  
    \put(60,75){\line(0,1){10}}
    \put(50,75){\line(1,0){10}}
    \put(50,85){\line(1,0){10}}
    \put(50,75){\line(0,1){10}}
    \put(40,75){\line(1,0){10}}
    \put(40,85){\line(1,0){10}}
    \put(40,75){\line(0,1){10}}
    \put(30,75){\line(1,0){10}}
    \put(30,85){\line(1,0){10}}
    \put(30,75){\line(0,1){10}}
    \put(20,75){\line(1,0){10}}
    \put(20,85){\line(1,0){10}}
    \put(20,75){\line(0,1){10}}
    \put(75,20){\line(0,1){10}}
    \put(85,20){\line(0,1){10}}
    \put(75,20){\line(1,0){10}}
    \put(75,30){\line(1,0){10}}
    \put(95,20){\line(0,1){10}}
    \put(85,20){\line(1,0){10}}
    \put(85,30){\line(1,0){10}}
    \put(75,30){\line(0,1){10}}
    \put(85,30){\line(0,1){10}}
    \put(75,30){\line(1,0){10}}
    \put(75,40){\line(1,0){10}}
    \put(95,30){\line(0,1){10}}
    \put(85,30){\line(1,0){10}}
    \put(85,40){\line(1,0){10}}
    \put(75,40){\line(0,1){10}}
    \put(85,40){\line(0,1){10}}
    \put(75,40){\line(1,0){10}}
    \put(75,50){\line(1,0){10}}
    \put(95,40){\line(0,1){10}}
    \put(85,40){\line(1,0){10}}
    \put(85,50){\line(1,0){10}}    
    
    \put(75,65){\line(0,1){10}}
    \put(85,65){\line(0,1){10}}
    \put(75,65){\line(1,0){10}}
    \put(75,75){\line(1,0){10}}
    \put(95,65){\line(0,1){10}}
    \put(85,65){\line(1,0){10}}
    \put(85,75){\line(1,0){10}}
    \put(75,75){\line(0,1){10}}
    \put(85,75){\line(0,1){10}}
    \put(75,75){\line(1,0){10}}
    \put(75,85){\line(1,0){10}}
    \put(100,20){\mbox{, $\;\varnothing$}}
    \put(72,25){\circle*{1}}
    \put(68,25){\circle*{1}}
    \put(64,25){\circle*{1}}
    \put(72,45){\circle*{1}}
    \put(68,45){\circle*{1}}
    \put(64,45){\circle*{1}}
    \put(72,70){\circle*{1}}
    \put(68,70){\circle*{1}}
    \put(64,70){\circle*{1}}
    \put(72,80){\circle*{1}}
    \put(68,80){\circle*{1}}
    \put(64,80){\circle*{1}}
    
    \put(25,54){\circle*{1}}
    \put(25,58){\circle*{1}}
    \put(25,62){\circle*{1}}
    \put(55,54){\circle*{1}}
    \put(55,58){\circle*{1}}
    \put(55,62){\circle*{1}}
    \put(80,54){\circle*{1}}
    \put(80,58){\circle*{1}}
    \put(80,62){\circle*{1}}
    \put(90,54){\circle*{1}}
    \put(90,58){\circle*{1}}
    \put(90,62){\circle*{1}}
    \put(53,5){\mbox{$m$}}
    \put(70,8){\vector(1,0){23}}
    \put(45,8){\vector(-1,0){23}}

    \put(5,49){\mbox{$n$}}
    \put(8,61){\vector(0,1){23}}
    \put(8,43){\vector(0,-1){23}}
  \end{picture}\right)
\end{equation*}
In the remaining integral the pole $P_1+(m-1)b+(n-1)b^{-1}$ will be inside all contours, while $P_2-Q$ stay outside. As a result, we restrict the set of all possible pairs of Young diagrams to the subset such that the first diagram contains $m\times n$ rectangle in it. From \eqref{Partition-function-N=2*-redefined} it is evident that any pair in the supplemental subset (i.e. the set of all pairs such that the first diagram does not contain $m\times n$ rectangle) does not have singularity at $P_1+(m-1)b+(n-1)b^{-1}\rightarrow P_2-Q$.  Of course there are dual poles when $P_2+(m-1)b+(n-1)b^{-1}$ collide with $P_1-Q$. Both of them correspond to
\begin{equation}\label{Z-poles}
    \Delta(P)\rightarrow\Delta_{m,n}.
\end{equation}
By doing simple algebra we get from \eqref{Zn-integral}
\begin{equation}
   \text{Res}\,\,\mathfrak{Z}_N(\alpha,\Delta)\Biggl|_{\Delta=\Delta_{m,n}}\hspace*{-15pt}=
     R_{m,n}(\alpha)\,\mathfrak{Z}_{N-mn}(\alpha,\Delta_{m,-n}),
\end{equation}
where $R_{m,n}(\alpha)$ is given by \eqref{Rmn}. One can show that poles \eqref{Z-poles} are the only singularities of the integral \eqref{Zn-integral} as a function of $P_1$ and $P_2$ (see more on this in appendix \ref{Z-poles-appendix}). That proves that the singular part of the partition function $Z_{\text{inst}}^{\mathcal{N}=2^{*}}(\varepsilon_1,\varepsilon_2,m,\vec{a})$ coinsides with the singular part of the conformal block \eqref{one-point-block-expansion}. 

Non-singular part of the partition function $Z_{\text{inst}}^{\mathcal{N}=2^{*}}(\varepsilon_1,\varepsilon_2,m,\vec{a})$ can be obtained in the limit $\Delta\rightarrow\infty$. From \eqref{Partition-function-N=2*-redefined} one gets
\begin{multline}\label{Z_N-limit}
    \mathfrak{Z}_N\xrightarrow[\Delta\rightarrow\infty]{}\zeta_N\bigl(\Delta(\alpha)\bigr)=
    \sum_{\vec{Y}}\prod_{\substack{s\in Y_1\\t\in Y_2}}
    \biggl(1-\frac{\Delta(\alpha)}{\Delta\bigl((-bH_{Y_1}(s)+b^{-1}(1+V_{Y_1}(s))\bigr)}\biggr)\times\\\times
    \biggl(1-\frac{\Delta(\alpha)}{\Delta\bigl((-bH_{Y_2}(t)+b^{-1}(1+V_{Y_2}(t))\bigr)}\biggr).
\end{multline}
It follows from \eqref{Z_N-limit} that
\begin{equation}
      1+\sum_{k=1}^{\infty}\zeta_k\bigl(\Delta(\alpha)\bigr)q^k=
      \left(1+\sum_{k=1}^{\infty}\xi_k\bigl(\Delta(\alpha)\bigr)q^k\right)^2, 
\end{equation}
where
\addtocounter{equation}{-1}
\begin{subequations}
\begin{equation}\label{Z_N-limit-root}
     \xi_N\bigl(\Delta(\alpha)\bigr)=
    \sum_{Y}\prod_{s\in Y}
    \biggl(1-\frac{\Delta(\alpha)}{\Delta\bigl((-bH_{Y}(s)+b^{-1}(1+V_{Y}(s))\bigr)}\biggr).
\end{equation}
\end{subequations}
In \eqref{Z_N-limit-root} the sum goes over all Young diagrams $Y$ with the total number of cells equal to $N$. It is not obvious from the explicit form \eqref{Z_N-limit-root} but $\xi_N\bigl(\Delta(\alpha)\bigr)$ does not depend on $b$. This statement can be proved using results of appendix \ref{Z-poles-appendix}. It is convenient to choose $b=\imath$. Then we have
\begin{equation}\label{Z_N-limit-root-i}
    \xi_N\bigl(\Delta(\alpha)\bigr)=
    \sum_{Y}\prod_{s\in Y}
    \biggl(1-\frac{\Delta(\alpha)}{\bigl(1+H_Y(s)+V_Y(s)\bigr)^2}\biggr).
\end{equation}
It was proved in \cite{Nekrasov:2003rj,han-2008} that
\begin{equation}
   1+\sum_{k=1}^{\infty}\xi_k\bigl(\Delta(\alpha)\bigr)q^k=
   \left(\frac{q^{\frac{1}{24}}}{\eta(\tau)}\right)^{1-\Delta(\alpha)}, 
\end{equation}
so that finally proves the relation \eqref{AGT-relation}.
\section{Seiberg-Witten prepotential}\label{SW}
It was argued in \cite{Nekrasov:2002qd,Nekrasov:2003rj} that the partition function \eqref{Partition-function-N=2*-def} has the following limit at $\varepsilon_1,\varepsilon_2\rightarrow0$ 
\begin{equation}
   Z_{\text{inst}}^{\mathcal{N}=2^{*}}(\varepsilon_1,\varepsilon_2,m,\vec{a})\rightarrow
   e^{\frac{1}{\varepsilon_1\varepsilon_2}\mathbb{F}(m,\vec{a}|q)},
\end{equation}
with $\mathbb{F}(m,\vec{a}|q)$ being the instanton part of the Seiberg-Witten prepotential \cite{Seiberg:1994aj}. Having in mind \eqref{parms-rescale} we set
\begin{equation}
  \Delta(\alpha)=-\frac{m^2}{\hbar^2}\quad\text{and}\quad\Delta(P)=-\frac{a^2}{\hbar^2},
\end{equation}
where $a=\frac{a_1-a_2}{2}$ and  consider the limit of the conformal block $\mathcal{H}_{\alpha}^{(\Delta)}(q)$ at $\hbar\rightarrow0$ \footnote{Conformal blocks $\mathcal{H}_{\alpha}^{(\Delta)}(q)$ and $\mathcal{F}_{\alpha}^{(\Delta)}(q)$ have the same limit since they differ from each other by finite factor (see \eqref{F-H-definition}).}
\begin{equation}\label{confblock-limit}
   \mathcal{H}_{\alpha}^{(\Delta)}(q)\xrightarrow[\hbar\rightarrow0]{}
    e^{\frac{1}{h^2}\mathbb{H}(m,a|q)}.
\end{equation}
The fact that the limit of $\mathcal{H}_{\alpha}^{(\Delta)}(q)$  has the form \eqref{confblock-limit} is rather non-trivial from its explicit definition.  Generally coefficients $H_L$ defined by \eqref{Hblock-series} have asymptotic at $\hbar\rightarrow0$
\begin{equation}
    H_L\sim\frac{1}{\hbar^{2L}}, 
\end{equation}
but  as stated in \eqref{confblock-limit}
\begin{equation}\label{Log-of-H}
   \log\left(\mathcal{H}_{\alpha}^{(\Delta)}(q)\right)=H_1q+\left(H_2-\frac{H_1^2}{2}\right)q^2+
   \left(H_3-H_1H_2+\frac{H_1^3}{3}\right)q^3+\dots
\end{equation}
behaves as $\hbar^{-2}$. It can be checked by explicit calculation that all unwanted terms in \eqref{Log-of-H} which have behavior $\sim\hbar^{-2k}$ with $k>2$ are cancelled. Moreover, function $\mathbb{H}(m,a|q)$ is $b$-independent and has the form
\begin{equation}\label{H-form}
    \mathbb{H}(m,a|q)=-\frac{m^4}{2a^2}\,H(u|q) 
\end{equation}
where $u=(\frac{m}{a})^2$. In \eqref{H-form} 
\begin{equation}
    H(u|q)=\sum_{k=1}^{\infty}p_k(u)q^k,
\end{equation}
with $p_k(u)$ being  polynomials of degree $2k-2$
\addtocounter{equation}{-1}
\begin{subequations}
\begin{equation}\label{polynomials}
   \begin{gathered}
     p_1(u)=1,\\
     p_2(u)=\frac{1}{32}\left(
     5\,{u}^{2}-48\,u+96
     \right),\\
     p_3(u)=\frac{1}{96}\left(
     9\,{u}^{4}-112\,{u}^{3}+480\,{u}^{2}-768\,u+384
     \right),\\\dots\dots\dots\dots
   \end{gathered}
\end{equation}
\end{subequations}
Polynomials \eqref{polynomials} are exactly the same as in \cite{Minahan:1997if}. The instanton part of the Seiberg-Witten prepotential $\mathbb{F}(m,\vec{a}|q)$ is related due to \eqref{AGT-relation} with $H(u|q)$ as
\begin{equation}\label{ZW-prep}
    \mathbb{F}(m,\vec{a}|q)=\frac{m^2}{12}\log(q)-2m^2\log(\eta(\tau))-\frac{m^4}{2a^2}\,H(u|q).
\end{equation}

In order to evaluate the limit of the conformal block at $\hbar\rightarrow0$ it is convenient to consider two-point correlation function with one degenerate field $V_{-\frac{b}{2}}$
\begin{equation}\label{Psi-definition}
    \langle V_{-\frac{b}{2}}(z)V_{\alpha}(0)\rangle_{\tau}=\left(\Theta_1(z)\right)^{\frac{b^2}{2}}
    \left(\eta(\tau)\right)^{2\Delta(\alpha)-1-2b^2}\,\Psi(z|\tau),
\end{equation}
where $\Theta_1(z)$ is elliptic theta-function. Field $V_{-\frac{b}{2}}$ is degenerate at the second level. As a consequence of this degeneracy function $\Psi(z|\tau)$ defined by \eqref{Psi-definition} satisfies differential equation of the second order (on a torus and other Riemann surfaces it was studied first time in \cite{Eguchi:1986sb}). In the limit $\hbar\rightarrow0$ one has
\begin{equation}\label{Schr-eq}
    \left(-\partial^2_z+\frac{b^2m^2}{\hbar^2}\wp(z)\right)\Psi(z|\tau)=\frac{2ib^2}{\pi}
    \,\partial_{\tau}\Psi(z|\tau),
\end{equation}
where $\wp(z)$ is Weierstra\ss ~ elliptic function (see appendix \ref{WP}). 
It is reasonable to look for the solution to \eqref{Schr-eq} in the form
\begin{equation}
   \Psi(z|\tau)=\exp\left(\frac{1}{\hbar^2}\,\mathcal{F}(q)+\frac{b}{\hbar}\,\mathcal{W}(z|q)+\dots\right).
\end{equation}
From  the definition of the function $\Psi(z|\tau)$ \eqref{Psi-definition} we expect that $\mathcal{F}(q)$ is
\begin{equation}\label{F-expectation}
    \mathcal{F}(q)=-a^2\log(q)+2m^2\log(\eta(\tau))-\frac{m^4}{2a^2}\,H(u|q).
\end{equation}
WKB approximation for the function $\mathcal{W}(z|q)$ gives
\begin{equation}\label{WKB-solution}
    \mathcal{W}(z|q)=\int_{z_0}^z\sqrt{E(q)+m^2\,\wp(z)}\,dz,
\end{equation}
where 
\addtocounter{equation}{-1}
\begin{subequations}
\begin{equation}\label{E-F-definition}
E(q)=4q\partial_q\mathcal{F}(q).
\end{equation}
\end{subequations}
The energy $E(q)$ can be fixed from the condition that the monodromy of the WKB solution \eqref{WKB-solution}  along the $A$ cycle of the torus is equal to $2i\pi a$
\begin{equation}\label{Energy-equation}
    \oint\limits_{A}\sqrt{E(q)+m^2\,\wp(z)}\,dz=2i\pi a.
\end{equation} 
For convenience we define $\mathcal{E}(q)=-\frac{1}{4 a^2}E(q)$
\begin{equation}\label{Energy-equation-improved}
    \oint\limits_{A}\sqrt{\mathcal{E}(q)-\frac{u}{4}\,\wp(z)}\,dz=\pi.
\end{equation} 
Equation \eqref{Energy-equation-improved} defines $\mathcal{E}(q)$ in parametric form. In particular, this form is very convenient for studying small $u$ expansion of the energy  $\mathcal{E}(q)$ 
\begin{equation}\label{E-expansion}
   \mathcal{E}(q)=1+\mathcal{E}_1(q)\,u+\mathcal{E}_2(q)\,u^2+\mathcal{E}_3(q)\,u^3+\dots
\end{equation}
Expanding equation \eqref{Energy-equation-improved} and using formulae from appendix \ref{WP} one can find coefficients  $\mathcal{E}_k(q)$
\begin{equation}
    \mathcal{E}_1(q)=\frac{g_1}{4},\quad\mathcal{E}_2(q)=\frac{g_2}{768}-\frac{g_1^2}{64},\quad
    \mathcal{E}_3(q)=\frac{g_3-g_1g_2}{5120}+\frac{g_1^3}{256},\quad\dots
\end{equation}
where $g_1$, $g_2$ and $g_3$ are given by \eqref{g23}-\eqref{g1}. Expansion \eqref{E-expansion} coinsides with the expansion suggested by \eqref{F-expectation} to higher orders in $u$. We note that within this approach we can obtain only the instanton part of the Seiberg-Witten prepotential\footnote{Eq. \eqref{SW-1} follows from \eqref{ZW-prep} and \eqref{F-expectation}.} 
\begin{equation}\label{SW-1}
   \mathbb{F}(m,\vec{a}|q)=\left(a^2+\frac{m^2}{12}\right)\log(q)-4m^2\log(\eta(\tau))+\mathcal{F}(q),
\end{equation}
where $\mathcal{F}(q)$ is defined as a solution to \eqref{Energy-equation},
while the perturbative part is the integration constant in \eqref{E-F-definition} which has to be fixed by other principles. We want to stress that in particular case when $u=4$ (at this point a charged BPS state is massless) equation \eqref{Energy-equation-improved} has a simple solution
\begin{equation}
    \mathcal{E}(q)=\wp\left(\frac{\pi}{2}\right)=
    8 q\partial_q\log\left(\frac{\eta(2\tau)}{\eta(\tau)}\right).
\end{equation}
We note also, that potential in Schr\"odinger equation \eqref{Schr-eq} is double periodic function and we can define dual quasi-momentum  as integral over $B$ cycle of the torus
\begin{equation}\label{Energy-equation-dual}
    \oint\limits_{B}\sqrt{E(q)+m^2\,\wp(z)}\,dz=2i\pi a_D,
\end{equation}  
which is the derivative of the total prepotential with respect to $a$ \cite{Seiberg:1994aj}.
\acknowledgments
We thank Eugeny Andriyash, Mikhail Bershtein, Sergei Lukyanov, Gregory Moore and Alexander Zamolodchikov for stimulating discussions and interest to this work.

This work was supported, in part, by cooperative CNRS-RFBR  grant 09-02-93106-CNRS. Work of A.~L. was supported  by DOE grant DE-FG02-96ER40949,  by RFBR  initiative interdisciplinary project grant 09-02-12446-OFI-m by Russian Ministry of Science and Technology under the Scientific Schools grant 3472.2008.2 and by RAS program "Elementary particles and the fundamental nuclear physics". The research of A.~L. was held within the framework of the federal program ”Scientific and Scientific-Pedagogical Personnel of Innovational Russia” on 2009-2013 (state contract No. P1339).
\appendix
\section{Structure of singularities of the integral \protect\eqref{Zn-integral}}\label{Z-poles-appendix}
In this appendix we consider pole structure of the integral \eqref{Zn-integral} in more details. First of all if one looks at the explicit expression \eqref{Partition-function-N=2*-redefined} one may have the impression that the integral \eqref{Zn-integral} has poles (even not necessary simple poles) at the points 
\begin{equation*}
    P_1-P_2=mb+nb^{-1}
\end{equation*}
with $m$ and $n$ being two arbitrary integer numbers restricted by the condition $|mn|\leq N$. Indeed,  contribution of the particular pair of Young diagrams to \eqref{Partition-function-N=2*-redefined} can have any of the above singularities, but the sum over all possible pairs does not. The only surviving singularities are the simple poles at $P_1-P_2=mb+nb^{-1}$ with eather $m$ and $n$ being both greater  than $1$ or being both less  than $-1$ with the condition $mn\leq N$. To prove it we consider more general integral
\begin{multline}\label{general-Young-integral}
   \Omega_N(P_1,P_2)=\frac{1}{N!}\left(\frac{Q}{2\pi i}\right)^N\times\\\times
   \oint\limits_{\mathcal{C}_1}\dots\oint\limits_{\mathcal{C}_N}
   \frac{F(x_1,\dots,x_N)}{\prod_{k}(x_k-P_1)(x_k-P_2)}\,
   \prod_{i<j}\frac{x_{ij}^2(x_{ij}^2-Q^2)}{(x_{ij}^2-b^2)(x_{ij}^2-b^{-2})}\,dx_1\dots dx_N,
\end{multline}
where the integration contours are exactly the same as in \eqref{Zn-integral} and $F(x_1,\dots,x_N)$ is some entire function\footnote{Without loss of generality we can assume that $F(x_1,\dots,x_N)$ is symmetric function of its variables.}. Integral \eqref{general-Young-integral} can expressed as a sum over pairs of Young diagrams with the total number of cells equal to $N$
\begin{equation}\label{general-Young-integral-answer}
    \Omega_N(P_1,P_2)=\sum_{\vec{Y}}\,g(\vec{Y})\,F(\vec{Y}),
\end{equation}
where by $F(\vec{Y})$ we denote the function $F(x_1,\dots,x_N)$ evaluated on a pair $\vec{Y}=(Y_1,Y_2)$. For example for the pair
\begin{equation*}
     \vec{Y}=(Y_1,Y_2)=\left(
   \begin{picture}(105,18)(15,20)
    \Thicklines
    \unitlength 1pt 
    \put(20,20){\line(0,1){10}}
    \put(30,20){\line(0,1){10}}
    \put(20,20){\line(1,0){10}}
    \put(20,30){\line(1,0){10}}
    \put(40,20){\line(0,1){10}}
    \put(30,20){\line(1,0){10}}
    \put(30,30){\line(1,0){10}}
    \put(50,20){\line(0,1){10}}
    \put(40,20){\line(1,0){10}}
    \put(40,30){\line(1,0){10}}
    \put(60,20){\line(0,1){10}}
    \put(50,20){\line(1,0){10}}
    \put(50,30){\line(1,0){10}}    
    \put(75,20){\line(0,1){10}}
    \put(85,20){\line(0,1){10}}
    \put(75,20){\line(1,0){10}}
    \put(75,30){\line(1,0){10}}
    \put(95,20){\line(0,1){10}}
    \put(85,20){\line(1,0){10}}
    \put(85,30){\line(1,0){10}}
    \put(100,20){\mbox{, $\;\varnothing$}}
    \put(72,25){\circle*{1}}
    \put(68,25){\circle*{1}}
    \put(64,25){\circle*{1}}
    \put(53,5){\mbox{$N$}}
    \put(70,8){\vector(1,0){23}}
    \put(45,8){\vector(-1,0){23}}
  \end{picture}\right)
\end{equation*}
one has
\begin{equation*}
     F(\vec{Y})\overset{\text{def}}{=}F(P_1,P_1+b,\dots,P_1+(N-1)b).
\end{equation*}
Rational function $g(\vec{Y})$ in \eqref{general-Young-integral-answer} is given by
\begin{equation}
     g(\vec{Y})=\prod_{i,j=1}^2\prod_{s\in Y_i}\frac{h_{ij}(s)}{E_{ij}(s)(Q-E_{ij}(s))},
\end{equation}
where $E_{ij}(s)$ is defined by \eqref{E_ij-definition} and
\begin{equation*}
   h_{ij}(s)=P_i-P_j+\alpha b+\beta b^{-1}\quad\text{for}\quad s=(\alpha,\beta).
\end{equation*}
We claim that $\Omega_N(P_1,P_2)$ is entire function of $P_1$ and $P_2$ (we prove this and even more general statement lately in this appendix). One can notice that if $F(x_1,\dots,x_N)$ is some polynomial of total degree less than $N$ then $\Omega_N(P_1,P_2)$ is identically zero.
Now let $F(x_1,\dots,x_N)$ has a pole at $x_k=\xi$ 
\begin{equation}
    F(x_1,\dots,x_N)\sim\frac{1}{\prod_k(x_k-\xi)},
\end{equation}
which is supposed to be outside the integration region in the integral \eqref{general-Young-integral}. In this case $\Omega_N(P_1,P_2)$ will have exactly the same form \eqref{general-Young-integral-answer}, but now because $F(x_1,\dots,x_N)$ has singularities itself $\Omega_N(P_1,P_2)$ will have simple poles at 
\begin{equation}\label{new-poles}
       P_k+mb+nb^{-1}=\xi. 
\end{equation}
In our case \eqref{Zn-integral} we have two singularities of the function $F(x_1,\dots,x_N)$
\begin{equation}
       \xi=P_1-Q\quad\text{and}\quad \xi=P_2-Q,
\end{equation}
and both of them are outside the integration region. That leads to the simple poles of \eqref{Zn-integral} in the points
\begin{equation}
       P_{ij}=mb+nb^{-1}\quad\text{for}\quad m,n>1\quad\text{and}\quad mn\leq N. 
\end{equation}
The corresponding residue can be calculated as explained in section \ref{Nekrasov-partition-funtion}. In principle, we also have poles related with $\alpha$ in the integral \eqref{Zn-integral}, but as it follows from \eqref{Partition-function-N=2*-redefined} they do not lead  to any singularities.

\begin{figure}
\psfrag{p1}{$P_k$}
\psfrag{x}{$x_j+\varepsilon_a$}
\psfrag{y}{$x_j-\varepsilon_a$}
	\centering
	\includegraphics[width=.3\textwidth]{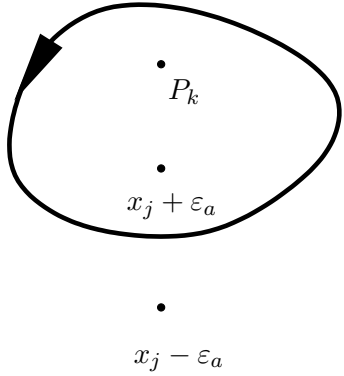}
        \caption{Integration contour $\mathcal{C}_i$ in  \protect\eqref{general-Young-integral-proof}.}
	\label{contour2}
\end{figure}

Now let us prove the statement announced above. Namely, function 
\begin{equation*}
       \Xi_N^{(F)}(P_1,\dots,P_n|\varepsilon_1,\dots,\varepsilon_p)=\Xi_N^{(F)}(P_k|\varepsilon_a)
\end{equation*}
defined by the integral\footnote{We note that integrals of the same type were studied in \cite{Moore:1998et,Moore:1997dj}.}
\begin{equation}\label{general-Young-integral-proof}
   \Xi_N^{(F)}(P_k|\varepsilon_a)=\frac{1}{N!}\left(\frac{1}{2\pi i}\right)^N
   \oint\limits_{\mathcal{C}_1}\dots\oint\limits_{\mathcal{C}_N}
   \frac{F(x_1,\dots,x_N)}{\prod_{k}\prod_p(x_k-P_p)}\,
   \prod_{i<j}\prod_a\frac{1}{(x_{ij}^2-\varepsilon_a^2)}\,dx_1\dots dx_N,
\end{equation}
where contour $\mathcal{C}_i$ goes as shown on fig \ref{contour2}, i.e. surrounds poles $P_k$ as well as $x_j+\varepsilon_a$ and leaves poles $x_j-\varepsilon_a$ outside is entire function of $P_k$ if $F(x_1,\dots,x_N)$ is some entire function of $x_1,\dots,x_N$.
Let us perform integration with respect to $x_N$. It is convenient to represent the contour $\mathcal{C}_N$ as
\begin{equation}
   \mathcal{C}_N=\mathcal{C}^{(0)}_N-\mathcal{C}^{(D)}_N,
\end{equation}
where $\mathcal{C}^{(0)}_N$ is the contour which surrounds all possible singularities and $\mathcal{C}^{(D)}_N$ the contour which surrounds the points $x_j-\varepsilon_a$. We don't worry to much about the integral over $\mathcal{C}_N^{(0)}$, because the result of integration will be entire function so no new singularities will appear after this integration. Integration over $\mathcal{C}^{(D)}_N$ is a trivial exercise leading to the new poles 
\begin{equation}
    x_j=P_k+\varepsilon_a,
\end{equation}
which by definition are inside the contours $\mathcal{C}_j$. Also, using  identity
\begin{equation*}
    \frac{1}{\varepsilon_a}\frac{1}{(x_{ij}-\varepsilon_a)^2-\varepsilon_b^2}+
    \frac{1}{\varepsilon_b}\frac{1}{(x_{ji}-\varepsilon_b)^2-\varepsilon_a^2}=
    \frac{\varepsilon_a+\varepsilon_b}{\varepsilon_a\varepsilon_b}\frac{1}{x_{ij}^2-(\varepsilon_a+\varepsilon_b)^2} 
\end{equation*}
one can show that other new poles will be in the points
\begin{equation}
    x_{ij}=\varepsilon_a+\varepsilon_b,
\end{equation}
and again by definition  the contour $\mathcal{C}_i$ surrounds $x_j+\varepsilon_a+\varepsilon_b$ but not $x_j-\varepsilon_a-\varepsilon_b$.
The result of integration over variable $x_N$ can be represented as
\begin{equation}
       \Xi_N^{(F)}(P_k|\varepsilon_a)=\Xi_{N-1}^{(\tilde{F})}(\tilde{P}_k|\tilde{\varepsilon}_a),
\end{equation}
where we defined
\begin{equation*}
  \begin{aligned}
      &\{\tilde{P}_k\}=\{P_1,\dots,P_n,P_1+\varepsilon_1,\dots,P_n+\varepsilon_1,\dots\},\\
      &\{\tilde{\varepsilon}_a\}=\{\varepsilon_1,\dots,\varepsilon_p,2\varepsilon_1,
       \varepsilon_1+\varepsilon_2,\dots\}
  \end{aligned}
\end{equation*}
and $\tilde{F}(x_1,\dots,x_{N-1})$ is some entire function which explicit expression is not important for us now. So as a result of integration over $x_N$ we have reduced our original $N$-dimensional integral \eqref{general-Young-integral-proof} to the $(N-1)$-dimensional integral of the same type. No singularities  (for example at $P_1\rightarrow P_2$) appear because they were inside our contour. We can repeat this procedure again and reduce it to the $(N-2)$-dimensional integral and so on. At any step we will get entire function of $P_1,\dots,P_n$. At the end we prove that $\Xi_N^{(F)}(P_k|\varepsilon_a)$ is entire function of $P_1,\dots,P_n$.
\section{Useful formulae}\label{WP}
In this appendix we collect some formulae for Weierstra\ss ~function.

The Weierstra\ss\, function can be expressed through the second logarithmic derivative of the theta-function $\Theta_1(z)$ as
\begin{equation}
    \wp(z)=\left(\frac{\Theta_1'(z)}{\Theta_1(z)}\right)^2-\frac{\Theta_1''(z)}{\Theta_1(z)}+
    \frac{1}{3}\frac{\Theta_1'''(0)}{\Theta_1'(0)}.
\end{equation}
This is double periodic function with periods $\pi$ and $\pi\tau$ which has the expansion at the origin
\begin{equation}
   \wp(z)=\frac{1}{z^2}+\frac{g_2}{20}\,z^2+\frac{g_3}{28}\,z^4+O(z^6), 
\end{equation}
where the invariants $g_2$ and $g_3$ enter in the differential equation satisfied by $\wp(z)$
\begin{equation}
    \wp'(z)^2=4\wp(z)^3-g_2\,\wp(z)-g_3, 
\end{equation}
and are given by
\begin{equation}\label{g23}
        \begin{aligned}
         &g_2=\frac{4}{3}\left(1+240\sum_{k=1}^{\infty}\sigma_3(k)q^k\right),\\
         &g_3=\frac{8}{27}\left(1-504\sum_{k=1}^{\infty}\sigma_5(k)q^k\right),
        \end{aligned}
\end{equation}
where $\sigma_n(k)$ is the divisor sigma function. For our purposes we define also
\begin{equation}\label{g1}
       g_1\overset{\text{def}}{=}\frac{1}{3}\frac{\Theta_1'''(0)}{\Theta_1'(0)}=
       -\frac{1}{3}\left(1-24\sum_{k=1}^{\infty}\sigma_1(k)q^k\right). 
\end{equation}
While considering small $u$-expansion of the equation \eqref{Energy-equation-improved} one has to evaluate the integrals
\begin{equation}
      f_n=\frac{1}{\pi}\oint\limits_{A}\wp^n(z)\,dz.
\end{equation}
One can notice that $\wp^n(z)$ can always be represented as a sum of even derivatives of the function $\wp(z)$ itself
\begin{equation}\label{WPn-expansion}
     \wp^n(z)=\sum_{k=1}^{n-1}c^{(n)}_k\wp^{(2k)}(z)+a_n\wp(z)+b_n.
\end{equation}
Coefficients $c^{(n)}_k$ as well as $a_n$ and $b_n$ can be found expanding both hand sides of \eqref{WPn-expansion} at the origin. Integrating over $A$-cycle one can drop all derivative terms in \eqref{WPn-expansion} due to periodicity. Finally, one gets 
\begin{equation}
      f_n=a_ng_1+b_n.
\end{equation}
Explicitly, first few integrals $f_n$ are
\begin{equation}
      \begin{aligned}
        &f_1=g_1,\qquad
        &&f_2=\frac{g_2}{12},
        &&f_3=\frac{g_3}{10}+\frac{3g_1g_2}{20},\\
        &f_4=\frac{5g_2^2}{336}+\frac{g_1g_3}{7}.\qquad
        &&f_5=\frac{g_2g_3}{30}+\frac{7g_1g_3}{240},\qquad
        &&f_6=\frac{g_3^2}{55}+\frac{15g_2^3}{4928}+\frac{87g_1g_2g_3}{1540}.
      \end{aligned}
\end{equation}
\bibliographystyle{MyStyle} 
\bibliography{MyBib}

\providecommand{\href}[2]{#2}\begingroup\raggedright\begin{thebibliography}{10}

\bibitem{Alday:2009aq}
L.~F. Alday, D.~Gaiotto, and Y.~Tachikawa, {\it {Liouville Correlation
  Functions from Four-dimensional Gauge Theories}},
  \href{http://xxx.lanl.gov/abs/0906.3219}{{\tt arXiv:0906.3219}}.

\bibitem{Nekrasov:2002qd}
N.~A. Nekrasov, {\it {Seiberg-Witten Prepotential From Instanton Counting}},
  {\em Adv. Theor. Math. Phys.} {\bf 7} (2004) 831--864,
  [\href{http://xxx.lanl.gov/abs/hep-th/0206161}{{\tt hep-th/0206161}}].

\bibitem{Mironov:2009qt}
A.~Mironov and A.~Morozov, {\it {The Power of Nekrasov Functions}},  {\em Phys.
  Lett.} {\bf B680} (2009) 188--194,
  [\href{http://xxx.lanl.gov/abs/0908.2190}{{\tt arXiv:0908.2190}}].

\bibitem{Mironov:2009qn}
A.~Mironov and A.~Morozov, {\it {Proving AGT relations in the large-c limit}},
  {\em Phys. Lett.} {\bf B682} (2009) 118--124,
  [\href{http://xxx.lanl.gov/abs/0909.3531}{{\tt arXiv:0909.3531}}].

\bibitem{Marshakov:2009kj}
A.~Marshakov, A.~Mironov, and A.~Morozov, {\it {Zamolodchikov asymptotic
  formula and instanton expansion in $N=2$ SUSY $N_f=2N_c$ QCD}},  {\em JHEP}
  {\bf 11} (2009) 048, [\href{http://xxx.lanl.gov/abs/0909.3338}{{\tt
  arXiv:0909.3338}}].

\bibitem{Wyllard:2009hg}
N.~Wyllard, {\it {$A_{N-1}$ conformal Toda field theory correlation functions
  from conformal $N=2$ $SU(N)$ quiver gauge theories}},  {\em JHEP} {\bf 11}
  (2009) 002, [\href{http://xxx.lanl.gov/abs/0907.2189}{{\tt
  arXiv:0907.2189}}].

\bibitem{Mironov:2009by}
A.~Mironov and A.~Morozov, {\it {On AGT relation in the case of $U(3)$}},  {\em
  Nucl. Phys.} {\bf B825} (2010) 1--37,
  [\href{http://xxx.lanl.gov/abs/0908.2569}{{\tt arXiv:0908.2569}}].

\bibitem{Gaiotto:2009we}
D.~Gaiotto, {\it {N=2 dualities}},
  \href{http://xxx.lanl.gov/abs/0904.2715}{{\tt arXiv:0904.2715}}.

\bibitem{Fucito:2004gi}
F.~Fucito, J.~F. Morales, and R.~Poghossian, {\it {Instantons on quivers and
  orientifolds}},  {\em JHEP} {\bf 10} (2004) 037,
  [\href{http://xxx.lanl.gov/abs/hep-th/0408090}{{\tt hep-th/0408090}}].

\bibitem{Belavin:1984vu}
A.~A. Belavin, A.~M. Polyakov, and A.~B. Zamolodchikov, {\it Infinite conformal
  symmetry in two-dimensional quantum field theory},  {\em Nucl. Phys.} {\bf
  B241} (1984) 333--380.

\bibitem{Zamolodchikov:1985ie}
{\relax Al}.~B. Zamolodchikov, {\it {Conformal symmetry in two-dimensions: an
  explicit reccurence formula for the conformal partial wave amplitude}},  {\em
  Commun. Math. Phys.} {\bf 96} (1984) 419--422.

\bibitem{Fateev:2009me}
V.~A. Fateev, A.~V. Litvinov, A.~Neveu, and E.~Onofri, {\it {Differential
  equation for four-point correlation function in Liouville field theory and
  elliptic four-point conformal blocks}},  {\em J. Phys. A} {\bf 42} (2009)
  304011, [\href{http://xxx.lanl.gov/abs/0902.1331}{{\tt arXiv:0902.1331}}].

\bibitem{Poghossian:2009mk}
R.~Poghossian, {\it {Recursion relations in CFT and N=2 SYM theory}},  {\em
  JHEP} {\bf 12} (2009) 038, [\href{http://xxx.lanl.gov/abs/0909.3412}{{\tt
  arXiv:0909.3412}}].

\bibitem{Hadasz:2009db}
L.~Hadasz, Z.~Jaskolski, and P.~Suchanek, {\it {Recursive representation of the
  torus 1-point conformal block}},
  \href{http://xxx.lanl.gov/abs/0911.2353}{{\tt arXiv:0911.2353}}.

\bibitem{Zamolodchikov:2003yb}
{\relax Al}.~B. Zamolodchikov, {\it {Higher equations of motion in Liouville
  field theory}},  {\em Int. J. Mod. Phys.} {\bf A19S2} (2004) 510--523,
  [\href{http://xxx.lanl.gov/abs/hep-th/0312279}{{\tt hep-th/0312279}}].

\bibitem{Nekrasov:2003rj}
N.~Nekrasov and A.~Okounkov, {\it {Seiberg-Witten theory and random
  partitions}},  \href{http://xxx.lanl.gov/abs/hep-th/0306238}{{\tt
  hep-th/0306238}}.

\bibitem{han-2008}
G.-N. Han, {\it {An explicit expansion formula for the powers of the Euler
  Product in terms of partition hook lengths}},
  \href{http://xxx.lanl.gov/abs/0804.1849}{{\tt arXiv:0804.1849}}.

\bibitem{Seiberg:1994aj}
N.~Seiberg and E.~Witten, {\it {Monopoles, duality and chiral symmetry breaking
  in N=2 supersymmetric QCD}},  {\em Nucl. Phys.} {\bf B431} (1994) 484--550,
  [\href{http://xxx.lanl.gov/abs/hep-th/9408099}{{\tt hep-th/9408099}}].

\bibitem{Minahan:1997if}
J.~A. Minahan, D.~Nemeschansky, and N.~P. Warner, {\it {Instanton expansions
  for mass deformed N = 4 super Yang- Mills theories}},  {\em Nucl. Phys.} {\bf
  B528} (1998) 109--132, [\href{http://xxx.lanl.gov/abs/hep-th/9710146}{{\tt
  hep-th/9710146}}].

\bibitem{Eguchi:1986sb}
T.~Eguchi and H.~Ooguri, {\it {Conformal and Current Algebras on General
  Riemann Surface}},  {\em Nucl. Phys.} {\bf B282} (1987) 308--328.

\bibitem{Moore:1998et}
G.~W. Moore, N.~Nekrasov, and S.~Shatashvili, {\it {D-particle bound states and
  generalized instantons}},  {\em Commun. Math. Phys.} {\bf 209} (2000) 77--95,
  [\href{http://xxx.lanl.gov/abs/hep-th/9803265}{{\tt hep-th/9803265}}].

\bibitem{Moore:1997dj}
G.~W. Moore, N.~Nekrasov, and S.~Shatashvili, {\it {Integrating over Higgs
  branches}},  {\em Commun. Math. Phys.} {\bf 209} (2000) 97--121,
  [\href{http://xxx.lanl.gov/abs/hep-th/9712241}{{\tt hep-th/9712241}}].

\end{thebibliography}\endgroup
\end{document}